%% file: main.tex
\documentclass[10pt,conference]{IEEEtran}
\IEEEoverridecommandlockouts
\usepackage{xspace}
\usepackage{caption}
\usepackage{pdfpages}
\usepackage{hyperref}
\usepackage{amsmath,amsfonts}
\usepackage{algorithmic}
\usepackage{graphicx}
\usepackage[utf8]{inputenc} 
\usepackage{textcomp}
\usepackage{listings}
\usepackage{xcolor, soul}
\usepackage[all]{nowidow}
\usepackage[inline]{enumitem}
\usepackage[scaled]{beramono}
\usepackage{multirow}
\usepackage{tikz,pifont}
\usepackage{graphicx} 
\usepackage{grffile}
\usepackage{pgfplots}
\usepackage{cleveref}
\usepackage{color, colortbl}
\usepackage[many]{tcolorbox}
\usepackage{pdflscape}
\usepackage{makecell}
\usepackage{pbox}
\usetikzlibrary{shadows}
\usepackage{balance}
\usepackage{ragged2e}
\usepackage[tight,footnotesize]{subfigure}
\usepackage{amssymb}
\usepackage{fontawesome}

\pgfplotsset{compat=1.8}
\usepgfplotslibrary{statistics}

\colorlet{lightgray}{gray!30}

\sethlcolor{lightgray}

\definecolor{mygreen}{rgb}{0,0.6,0}
\definecolor{mygray}{rgb}{0.95,0.95,0.95}
\definecolor{myred}{rgb}{0.5,0,0}

\definecolor{Gray}{gray}{0.9}

\newcommand\untick{\ding{54}}

\newcommand\edited{\ding{45}}

\input{commands}

\begin{document}
	




\title{Is this piece of source code written by \CG?} 

\title{Tell me if this piece of code is written by \CG} 

\title{Tell me if this piece of code was written by \CG} 

\title{Tell me \GS: Is this piece of code written by \CG?} 

\title{\GS, is this piece of code written by \CG?} 

\title{Tell me \GS, is this piece of code \\ written by AI?} 

\title{Is this piece of code written by \CG?} 

\title{Is this piece of code written by \CG?\\ A source code detector using \CB} 

\title{Is this snippet written by \CG?\\ A source code detector using \CB} 

\title{Is this snippet written by \CG?\\ Detecting source code provenance using \CB} 

\title{Is this Snippet Written by \CG? An Empirical Study with a \CB-Based Classifier} 





\author{\IEEEauthorblockN{Phuong T. Nguyen}
\IEEEauthorblockA{\textit{Universit\`a degli studi dell'Aquila, Italy} \\
phuong.nguyen@univaq.it} 
\and
\IEEEauthorblockN{Juri Di Rocco}
\IEEEauthorblockA{\textit{Universit\`a degli studi dell'Aquila, Italy} \\
	juri.dirocco@univaq.it}
\and
\IEEEauthorblockN{Claudio Di Sipio}
\IEEEauthorblockA{\textit{Universit\`a degli studi dell'Aquila, Italy} \\
	claudio.disipio@graduate.univaq.it}
\and
\IEEEauthorblockN{Riccardo Rubei}
\IEEEauthorblockA{\textit{Universit\`a degli studi dell'Aquila, Italy} \\
	riccardo.rubei@graduate.univaq.it}
\and
\IEEEauthorblockN{Davide Di Ruscio}
\IEEEauthorblockA{\textit{Universit\`a degli studi dell'Aquila, Italy} \\
	davide.diruscio@univaq.it}
\and
\IEEEauthorblockN{Massimiliano Di Penta} 
\IEEEauthorblockA{\textit{Universit\`a degli studi del Sannio, Italy} \\
dipenta@unisannio.it}
}


\maketitle

\begin{abstract}
Since its launch in November 2022, \CG has gained popularity among users, especially programmers who use it as a tool to solve development problems. However, while offering a practical solution to programming problems, \CG should be mainly used as a supporting tool (e.g., in software education) rather than as a replacement for the human being. Thus, detecting automatically generated source code by \CG is necessary, and tools for identifying AI-generated content may need to be adapted to work effectively with source code.
This paper presents an empirical study to investigate the feasibility of automated identification of AI-generated code snippets, and the factors that influence this ability. To this end, we propose a novel approach called \GS, which builds on top of \CB to detect source code written by AI. The results show that \GS can accurately classify whether code is human-written or AI-generated, and outperforms two baselines, GPTZero and OpenAI Text Classifier. Also, the study shows how similar training data or a classification context with paired snippets helps to boost classification performances.


\end{abstract}


\thispagestyle{plain}

\section{Introduction}
\label{sec:Introduction}

\input{src/Introduction}

\section{Background and motivations} 
\label{sec:Background}
\input{src/Background}

\section{Infrastructure for the empirical study}
\label{sec:Approach}

\input{src/Approach}

\section{Empirical study design and methodology} 
\label{sec:Evaluation}
\input{src/Evaluation}

\section{Results}
\label{sec:Results}
\input{src/Results}

\section{Discussion}
\label{sec:Discussion}
\input{src/Discussion}

\section{Related work}
\label{sec:RelatedWork}
\input{src/RelatedWork}

\section{Conclusion and future work}
\label{sec:Conclusion}
\input{src/Conclusion}

\balance
\bibliographystyle{IEEEtranS}
\bibliography{IEEEabrv,main}

\end{document}

%% file: commands.tex
\usepackage{amssymb}

\newtcolorbox{shadedbox}{
	drop shadow southeast,
	breakable,
	enhanced jigsaw,
	colback=white,
	boxrule=0.80pt,
	left=0.3em,
	right=0.3em,
	top=0.1em,
	bottom=0.05em
}

\newcommand*{\ie}{i.e.,\@\xspace}
\newcommand*{\eg}{e.g.,\@\xspace}

\newcommand*{\GH}{GitHub\@\xspace}

\newcommand*{\CG}{ChatGPT\@\xspace}
\newcommand*{\GS}{GPTSniffer\@\xspace}
\newcommand*{\CB}{CodeBERT\@\xspace}
\newcommand*{\GZ}{GPTZero\@\xspace}
\newcommand*{\OT}{OpenAI Text Classifier\@\xspace}

\newcommand*{\numHumanSnippets}{601\@\xspace}
\newcommand*{\numGPTSnippets}{609\@\xspace}
\newcommand*{\DAlfa}{\textit{D}$_\alpha$\@\xspace}
\newcommand*{\DBeta}{\textit{D}$_\beta$\@\xspace}

\newcommand*{\numConf}{eight\@\xspace}

\newcommand{\api}[1]{\texttt{\hl{\small #1}}}

\makeatletter
\newcommand*{\etc}{%
	\@ifnextchar{.}%
	{etc}%
	{etc.\@\xspace}%
}
\makeatother
\newcommand*{\etal}{\emph{et~al.}\@\xspace}

\newcommand{\rqfirst}{\textbf{RQ$_1$}: \emph{How do the input data and the preprocessing settings impact on the \GS prediction performance?}}
\newcommand{\rqsecond}{\textbf{RQ$_2$}: \emph{To which extent can \GS detect \CG-generated source code on the paired dataset under different preprocessing settings?}}
\newcommand{\rqthird}{\textbf{RQ$_3$}: \emph{How does \GS compare with \GZ and \OT in recognizing \CG-generated source code?}}

\definecolor{verylightgray}{gray}{0.99}
\definecolor{lightgray}{gray}{0.92}

\definecolor{mygreen}{rgb}{0,0.6,0}
\definecolor{mygray}{rgb}{0.95,0.95,0.95}
\definecolor{myred}{rgb}{0.5,0,0}

\lstdefinestyle{JavaStyle} {
	backgroundcolor=\color{verylightgray},   
	commentstyle=\color{mygreen}, 
	breakatwhitespace=false,
	keywordstyle=\color{violet},
	language=Java,
	stringstyle=\color{blue},
	basicstyle=\scriptsize,
	showstringspaces=false
}

\lstdefinestyle{searchstringstyle}{
	basicstyle=\ttfamily\scriptsize,
	captionpos=t,                    
	numbers=none,                    
	numbersep=5pt,                  
	showspaces=false,                
	showstringspaces=false,
	showtabs=false,                  
	tabsize=2,
	frame=single
}

%% file: src/Introduction.tex
ChatGPT\footnote{GPT stands for Generative Pre-trained Transformer.} \cite{chatgpt} is a generative Artificial Intelligence (AI) tool, able to produce convincingly human answers to queries from users. Since its public release on November 30, 2022, \CG has attracted the attention of both expert- and non expert users worldwide, reaching one million users only five days after the launching.  

One of the areas in which \CG appears to be particularly promising, is its ability to support developers in a variety of tasks, that range from writing source code that fulfills a given (natural language) specification, to creating a software architecture/design, generating tests, or fixing a bug. 

Leveraging \CG---as well as some previously-existing AI-based code generation tools such as  \GH Copilot \cite{copilot}, OpenAI Codex \cite{codex}, or Tabnine \cite{tabnine}---to get recommendations for source code solutions is becoming very popular among developers. This does not happen without risks, as it has been shown that generative models could provide vulnerable code \cite{chatgptlaw,Pearce_Ahmad_Tan_Dolan-Gavitt_Karri_2021}, and, also, there is a wide yet controversial discussion on possible copyright and licensing infringements \cite{copilotcopyright,copilotLicense}.

Moreover, when \CG or other code generators are used by students during their learning processes, issues on risks and benefits arise, and this has triggered quite some discussion among educators. On the positive side,
code snippets generated by \CG provide students with a practical way to complete their assignments. 
At the same time, one major risk is that students would not  develop some essential skills that can be acquired only through self-learning, \eg critical thinking and problem-solving.  Moreover, handing in  code written by \CG without additional work can be considered as a form of fraud. 
Such behaviors trigger ethical concerns, as students have their work done without actually performing their own research. As a consequence, some universities have regulated, limited, or even banned the use of \CG.

As software engineering researchers, on the one hand, we need to promote
the democratized use of AI tools to facilitate daily programming tasks.  On the other hand, we believe that it is necessary
to recognize whether a source code element has been written by an AI for various reasons, and, in particular \emph{(i)} from the professional development side, dealing with security and legal problems; and \emph{(ii)} from the educational side, coping with cheating and plagiarism.


Recently, tools such as \GZ \cite{gptzero} and \OT \cite{ot} have been developed to 
 automatically recognize if a text is written by OpenAI technologies.
 Unfortunately, we noticed, by some attempts, that such tools are not necessarily good at  distinguishing between source code written by humans and AI. 
We conjecture that the underpinning engine has been trained on natural language text, rather than source code. This makes it necessary to train specific classifiers aimed at identifying AI-generated code.

\begin{figure*}[t!]
	\centering
	\begin{tabular}{c c }	
		\subfigure[Human-written code]{\label{fig:CodeHuman}
			\includegraphics[width=0.37\linewidth]{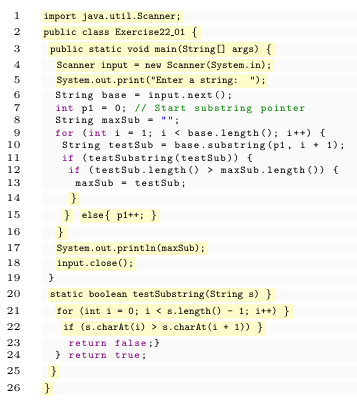}}	&
		\subfigure[\CG-generated code]{\label{fig:CodeChatGPT}\includegraphics[width=0.42\linewidth]{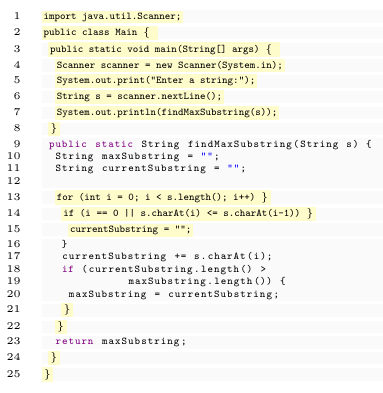}} 
	\end{tabular} 
	\vspace{-.2cm}
	\caption{Two code snippets to display the maximum consecutive increasingly ordered substring.} 	
	\label{fig:MotivatingExample}
	\vspace{-.4cm}
\end{figure*}

This paper presents an empirical study to investigate the extent to which it is possible to automatically detect whether a code snippet is written by \CG or humans, as well as the factors that can influence this ability. To achieve this, we present \GS--a machine learning solution to determine whether a piece of source code has been generated by \CG. The classification engine is based on \CB~\cite{feng-etal-2020-codebert}, a pre-trained model built on top of a code search dataset, \ie CodeSearchNet~\cite{DBLP:journals/corr/abs-1909-09436}. To the best of our knowledge, there is no specific approach able to identify whether source code has been generated by AI.

We evaluated \GS on two datasets collected from \GH and \CG. In the evaluation, we studied how characteristics of the training and test, and preprocessing steps impact the prediction performance. Also, we empirically compare \GS with \GZ and \OT.
The experimental results reveal interesting outcomes, while \GS cannot work well given that the training data and testing data are collected from completely independent sources, it obtains a perfect prediction by most of the configurations, where there are pairwise relationships between code written by humans and generated by \CG.
	

The main contributions of our work are the following ones: 

\begin{itemize}
	\item A novel approach--named \GS--to the recognition of source code generated by \CG.
	\item An empirical evaluation and comparison with state-of-the-art baselines, \GZ and \OT.
	\item The tool developed and the datasets curated through this work are made available to allow for future research \cite{replication}.
\end{itemize}

\textbf{Paper Structure.}  Section~\ref{sec:Background} provides a motivating example, and the proposed approach is described in Section~\ref{sec:Approach}.  
Section~\ref{sec:Evaluation} presents 
the materials and methods used to conduct an empirical evaluation on the proposed approach. Afterwards, Section~\ref{sec:Results} reports and analyzes the experimental results. We have some discussion and highlight the threats to validity in Section~\ref{sec:Discussion} The related work is reviewed in Section~\ref{sec:RelatedWork}, and the paper is concluded in Section~\ref{sec:Conclusion}.

%% file: src/Background.tex


As outlined in the introduction, concerns related to security, copyright/licensing infringement, or education ethics make particularly important  to identify whether a source code has been generated by and AI.

In principle, some solutions to cope with this problem exist. For example,
\GZ is one of the existing systems designed to automatically detect text generated by OpenAI technologies. However, by testing \GZ on source code, we notice that the outcome is far from satisfactory, suggesting how a well-defined text classifier fails to detect the origin of source code. 

Fig.~\ref{fig:MotivatingExample} shows an example with two code snippets, which are implemented exactly for the same purpose, \ie displaying the maximum consecutive increasingly ordered substring.\footnote{The original snippet is available online: \url{https://bit.ly/3MZCDWy}} However, one of them is written by humans (Fig.~\ref{fig:CodeHuman}), and the other one is generated by \CG (Fig.~\ref{fig:CodeChatGPT}). 
The snippets look pretty standard, \ie they use common API calls, such as \api{chatAt()}, \api{substring()}, or the \api{java.util.Scanner} package. Essentially, it is not easy to spot any concrete sign that can be used to 
recognize the source code's origin. 

We fed the code in Fig.~\ref{fig:CodeHuman} and Fig.~\ref{fig:CodeChatGPT} to \GZ, one by one, and asked for identification. Surprisingly, the platform gave the same conclusion for both snippets, \ie ``\emph{Your text is likely to be written entirely by a human}.'' This means that the system wrongly classifies the second snippet. Moreover, \GZ also added a remark, saying that: 
``\emph{Sentences highlighted are more likely to be written by AI}.'' Such sentences, \ie lines of code, are marked using yellow in Fig.~\ref{fig:MotivatingExample}. By comparing the designated parts in both snippets, we see that \GZ evaluates many common code lines as written by AI, \eg \api{public\; static\; void\; main(String[]\; args)} (Line 3) or \api{System.out.print(``Enter\; a\; string:'');} (Line 5). This is interesting as these lines can be written by both humans and \CG. Moreover, affirming that the snippets are ``\emph{written \textbf{entirely} by a human},'' while still highlighting lines that ``\emph{are more likely to be written by AI},'' is somewhat contradictory, rendering the classification result even more confusing.


By further testing \GZ with more code from humans and \CG, we witnessed similar outcomes.
One likely explanation is that, while the underlying GPT model of \GZ has been trained on a large corpus of text from the Internet (including source code too), it has not been specifically fine-tuned for source code.
Altogether, we see room for improvement, \ie 
the pattern in which commands are written, or the way comments are generated, are among distinguishable features that can be used to detect the origin of a snippet. This motivated us to investigate how well specifically-trained models can effectively recognize AI-generated source code, as it is described in the rest of the paper.

%% file: src/Approach.tex
This section describes \GS, the conceived tool to identify \CG-generated source code. It is worth noting that our aim here is not to propose a novel classification approach, but rather, to create an infrastructure for our empirical investigation, leveraging state-of-the-art pre-trained models.  Being built on top of \CB,  \GS inherits the well-defined technical foundation from the pre-trained model, attempting to achieve an ideal classification outcome for source code written in different languages.

\begin{figure}[t!]
	\centering
	\vspace{-.25cm}
	\includegraphics[width=0.45\textwidth]{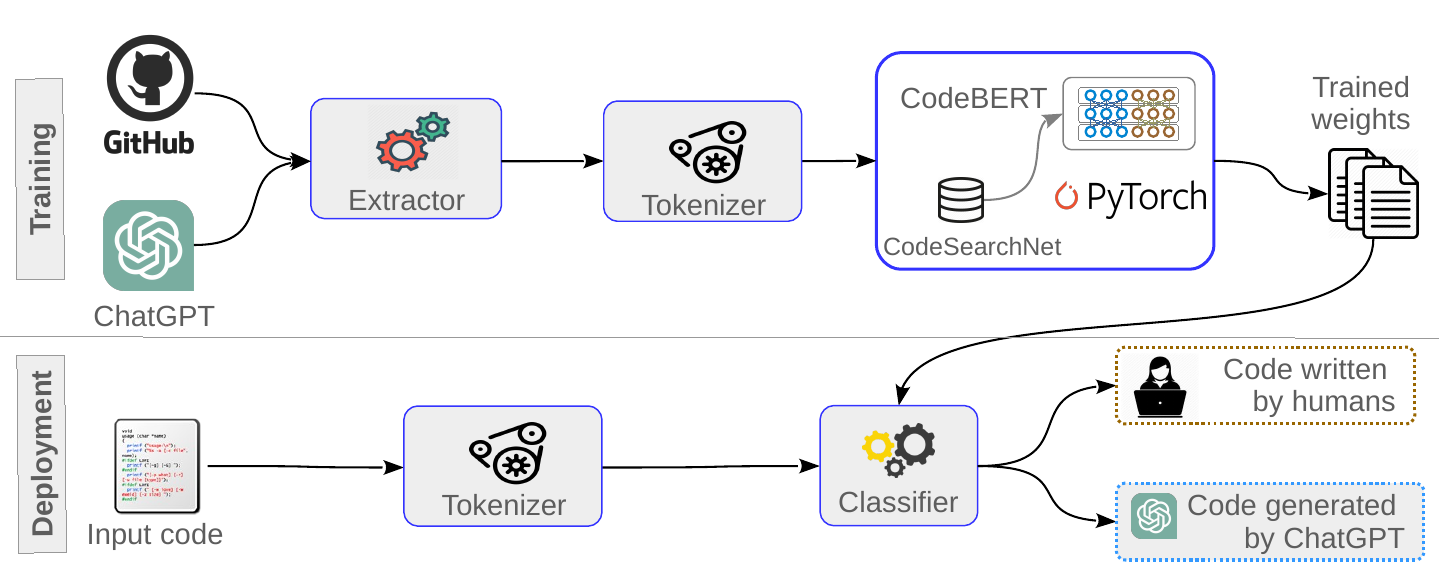}
	\caption{System components.}
	\vspace{-.2cm} 
	\label{fig:System}
	\vspace{-.4cm} 
\end{figure}

As shown in Fig.~\ref{fig:System}, the \GS architecture consists of three main components, \ie \emph{Extractor}, \emph{Tokenizer}, and \emph{Classifier}. To train the classification engine, input data is collected from two data sources, \ie \GH and \CG: while the former is a huge store of human-written code, the latter provides code generated by AI. Through the \emph{Extractor} component, the input data is then undergone different preprocessing steps to enrich the training corpus.  
\emph{Tokenizer} is employed to encode data for providing input to the \emph{Classifier} component, which performs the training to yield the final model. Such a model can then be used to perform a prediction for unseen code snippets. The \GS components are described in the following subsections.



%

\subsection{Extractor}

Empirical studies have shown that source code features such as package names, class names, code comments, or import directives are the unique features to identify the so-called \emph{coding style}~\cite{bosu2015characteristics,10.1145/3510003.3510181,DBLP:conf/wcre/OguraMHK18}. 
We conjecture that style-related features can be an effective means to distinguish snippets written by humans from those generated by \CG. 

The \emph{Extractor} component collects and prepares suitable data to train \GS. The ultimate aim is to create different derivations of the original code snippets, allowing the classifier to learn from diverse coding styles. \emph{Extractor} implements a set of rewriting rules, defined by adopting regular expressions. Given that an artifact, \eg imports, package names, or code comments, matches the regular expression, it can either be removed, or replaced with the one that resembles a certain coding style. 
 

\subsection{Tokenizer}

\begin{figure}[t]
	\centering
	\includegraphics[width=0.38\textwidth]{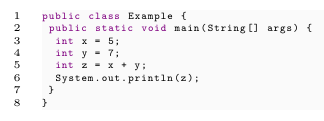}
	\caption{Example input code.}
	\vspace{-.2cm} 
	\label{fig:CodeBERT_tokenization}
	\vspace{-.3cm} 
\end{figure}

Once preprocessed by \emph{Extractor}, the source code is provided as input to the \emph{Tokenizer} component, which transforms the code into a proper format that can then be consumed by \CB. In particular, the input code is split into independent units called tokens, and padded with signaling tokens to separate the snippet from others. For example, the  class in Fig.~\ref{fig:CodeBERT_tokenization} is transformed by the \emph{Tokenizer} into the following sequence: \api{BOS,\; public,\; class,\; Example,\; \{,\; public,\; static,\; void,\; main,\; (,\; String,\; [,\; ],\; args,\; ),\; \{,\; int,\; x,\; =,\; 5,\; ;,\; int,\; y,\; =,\; 7,\; ;,\; int,\; z,\; =,\; x,\; +,\; y,\; ;,\; System,\; .,\; out,\; .,\; println,\; (,\; z,\; ),\; ;,\; \},\; \},\; EOS}. BOS and EOS are the two special tokens to signal the beginning and end of the sequence. The resulting sequence is then fed as input to the classification to perform the training and prediction.

\begin{figure*}[t!]
	\centering
	\vspace{-.1cm}
	\includegraphics[width=0.80\textwidth]{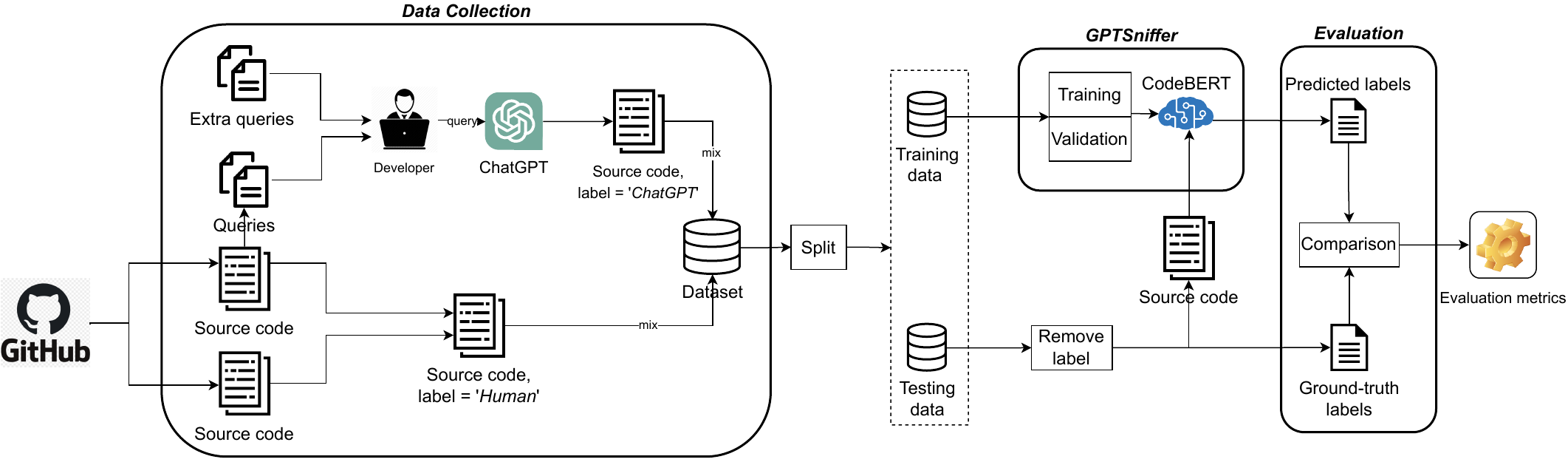}
	\caption{The evaluation process.}
	\vspace{-.4cm} 
	\label{fig:GPTSnifferWorkflow}
	\vspace{-.1cm} 
\end{figure*}

\subsection{Classifier}


\CB has been pre-trained on CodeSearchNet~\cite{DBLP:journals/corr/abs-1909-09436}, a code search dataset with more than 2M bimodal code-documentation pairs and 6.4M unimodal code snippets written in different languages, including Java, and Python. \emph{Classifier} is built on top of \CB,\footnote{We make use of the \CB pre-trained model provided by Huggingface (\url{https://huggingface.co/microsoft/codebert-base})} and run with Pytorch. In this respect, it inherits the well-founded technical features from the original model. Starting from the sequence of tokens generated by \emph{Tokenizer}, \emph{Classifier} uses a series of encoding layers to transform it into a fixed-length vector representation. Each encoding layer performs a series of computations on the input sequence to generate a new sequence of vectors that captures different aspects of the input's context.

%% file: src/Evaluation.tex



This section describes the empirical evaluation to study the \GS ability to detect \CG-generated snippets, and investigate the factors that impact on such ability. 


%
%
%
%

\subsection{Research Questions}

We study the performance of \GS 
by answering the following research questions:

\begin{itemize}
	\item \rqfirst~Using two datasets collected from \GH and \CG,  we conducted a series of experiments to identify the characteristics of the training data that can influence the accuracy of \GS, also under different preprocessing configurations.
	\item \rqsecond~In this case we put \GS under a particularly favorable scenario, \ie the presence of paired snippets (human vs. AI) in the training data, and investigated how \GS would perform in such a scenario under different code preprocessing configurations.
	\item \rqthird~Using a set of common queries, we compare \GS with \GZ \cite{gptzero} and \OT, \cite{ot} two state-of-the-art systems for identifying whether a text is generated by AI, including \CG, or written by humans. 
\end{itemize}



%



\subsection{Data Collection} \label{sec:DatCollection}

To simulate real-world scenarios, we collected data from different sources, as \GS is expected to detect code written by different developers. Also, we considered source code obtained by querying \CG under different conditions. The retrieval was performed following the process depicted in Fig.~\ref{fig:GPTSnifferWorkflow}, \ie we conducted two separate phases to obtain both unpaired and paired snippets, explained as follows. 

\subsubsection{Unpaired Snippets ($\mathcal{U}$)} \label{sec:IndependentSnippets}

To test the generalizability of \GS, we generated a set of additional queries to fetch code from \CG. 
Such queries cover a wide range of tasks, from simple to complex ones, aiming to study the usefulness of \GS. 
An example query is as follows: ``\emph{Can you write a Java program to implement the binary search algorithm?}'' The final corpus consists of \textbf{137 snippets} summarized in Table~\ref{tab:UnpairedData}. 

\begin{table}[h!]
	\vspace{-.1cm}
	\centering
	\small
	\footnotesize
	\caption{Summary on the additional queries.}  
	\vspace{-.2cm}
	\begin{tabular}{|p{1.2cm}|p{5cm}|p{1.2cm}|}    \hline  
		\textbf{Domain} & \textbf{Description} & \textbf{\# snippets}  \\ \hline 
		Algorithms & Feed-forward neural networks, convolutional neural networks, graph neural networks, Boyer Moore algorithm, Dijkstra algorithm, greatest common division, Levenshtein, logistic regression, matrix multiplication, Rocchio algorithm, SVM  & 35 \\ \hline 
		Files \& Folders & Append text, concatenate file, Excel file, read file, read file list, write file, copy file, get modified time, MP3 file &  25 \\ \hline 
		Networks & Email, HTTP client/server, FTP client/server, chat client/server  & 20  \\ \hline 
		Search \& Sort & Binary search, exponential search, sequential search, breadth-first search, depth-first search, linear search, bubble sort, merge sort &  22 \\ \hline
		Strings \& Arrays & Array blocking issue, compare two strings, delete word, dequeue, common elements, minimum element  &  15 \\ \hline
		Others & Binary tree, collaborative filtering, content-based filtering, hash table, lunar calendar, password checker, quadratic equation &  20 \\ \hline 
		 &  \textbf{Total} & \textbf{137}  \\ \hline
	\end{tabular}	
	\label{tab:UnpairedData}
\end{table}

Moreover, we independently collected \textbf{137 human-written snippets} being most starred from \GH Gist \cite{gist}. 
Unlike the paired snippets---that will be described in Section~\ref{sec:PairwiseSnippets}, the unpaired \CG-generated and human-written snippets are not related to each other. This simulates real use cases, where data is supposed to be collected from various repositories, and there exists no pairwise relationship between the snippets.



%


\subsubsection{Paired Snippets ($\mathcal{P}$)} \label{sec:PairwiseSnippets}

\begin{figure}[h!]
	\centering
	\vspace{-.2cm}
	\includegraphics[width=0.50\textwidth]{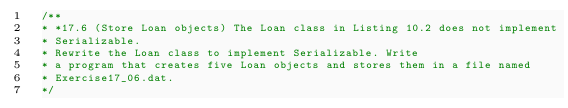}
	\caption{Example of a code assignment.}
	\label{fig:Assignment}
	\vspace{-.1cm} 
\end{figure}

We consider problem implementations from a book on Java programming~\cite{liang2003introduction}. Such a book has a supporting \GH repository \cite{introjavarepo}, storing the proposed solutions to the end-of-chapter exercises. We suppose that these snippets might have been used as training data for \CG, though we have no trace of that. 
Each code snippet is associated with a task assignment, placed at the beginning of the snippet as a source code comment. An example of such an assignment is shown in Fig.~\ref{fig:Assignment}.  

By manually scrutinizing them, we noticed that not all of the snippets are eligible for  
our experiments, as there are many of them containing only the assignment, without any code.  
Thus, these were not selected for the experiments. Eventually, we obtained a corpus containing \textbf{\numHumanSnippets human-written snippets}. 

\begin{figure}[h!]
	\centering
	\vspace{-.15cm}
	\includegraphics[width=0.44\textwidth]{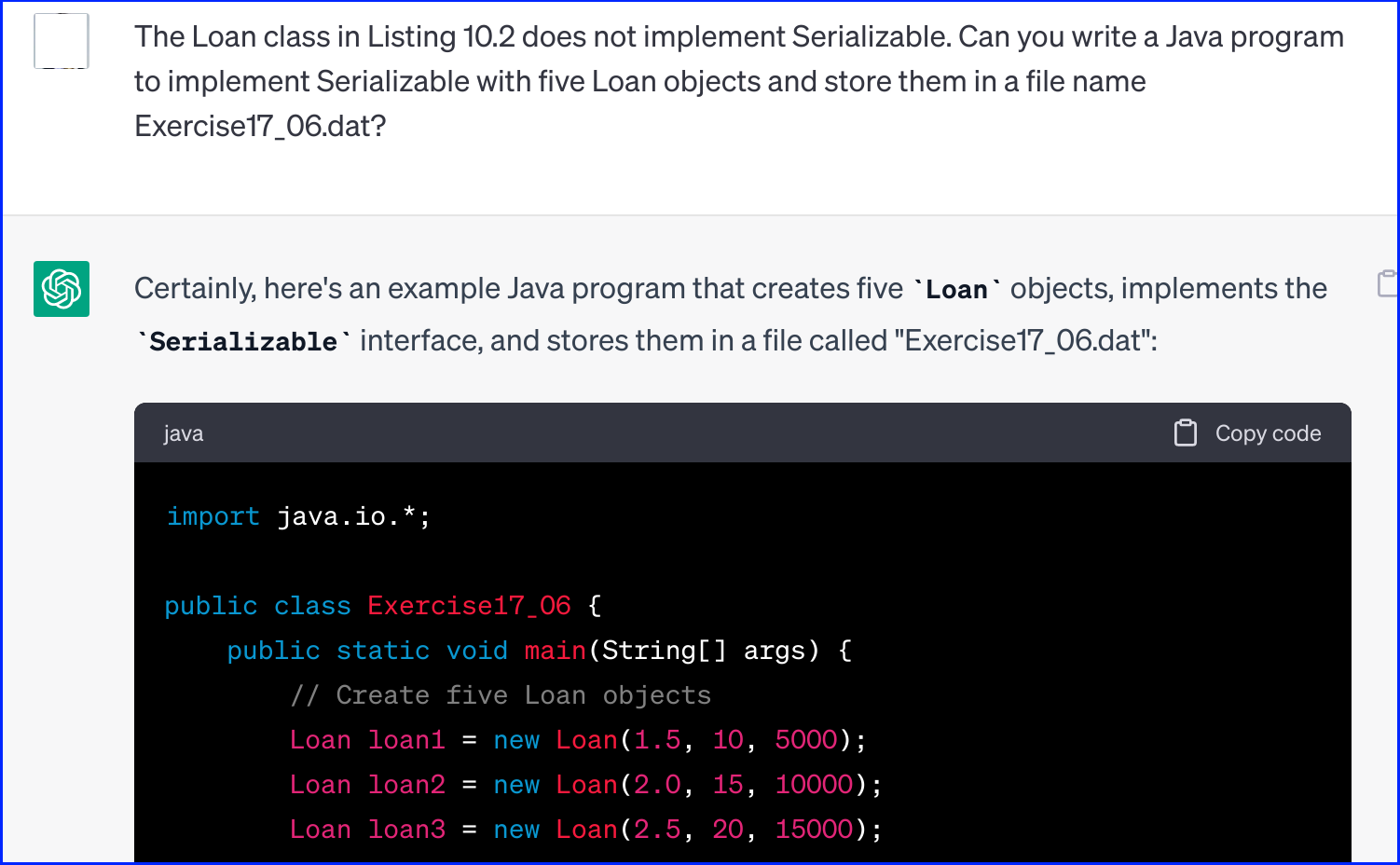}
	\caption{Interacting with \CG to retrieve source code.}
	\vspace{-.2cm} 
	\label{fig:CodeRetrieval}
	\vspace{-.1cm} 
\end{figure}

Starting from these human-written snippets, we extracted the task assignments and used them as queries. The queries were then split among the co-authors of this paper, that directly interacted with \CG to retrieve generated solutions to the assignments. Fig.~\ref{fig:CodeRetrieval} shows an example of interacting with \CG for the corresponding query transformed from the task assignment in Fig.~\ref{fig:Assignment}. After this step, we got a corpus of \textbf{\numGPTSnippets \CG-generated snippets}. Although the queries have been extracted from \numHumanSnippets human-written snippets, we got few \CG implementations more as some of them are split among different snippets.

By counting the number of lines of code (LOC) for all the snippets collected from humans and \CG by unpaired $\mathcal{U}$, and paired $\mathcal{P}$ snippets, we see that most of the snippets have a small LOC, \ie lower than 80. Only a few of them are longer than 100 LOC. As for the code written by humans, there are some considerably long snippets, with up to 1,200 LOC.

Once we obtained the two initial sets of snippets, 
we populated two datasets as shown in Table~\ref{tab:Datasets}. 

\begin{itemize}
	\item Dataset \DAlfa: We shuffled all the snippets in Sections~\ref{sec:IndependentSnippets} and~\ref{sec:PairwiseSnippets}, and split again to distribute the snippets coming from different sources into balanced parts. This aims to simulate real-world scenarios, where either human-written or \CG code can be collected in different ways. 
	The resulting \emph{mixed dataset} \DAlfa consists of 1,484 human-written and \CG-generated snippets. 
	\item Dataset \DBeta: We considered only the paired snippets related to the book's implementation in Section~\ref{sec:PairwiseSnippets}, resulting in the \emph{paired dataset} \DBeta with 1,210 snippets.
\end{itemize}

Note that \DAlfa and \DBeta are not independent, as \DAlfa is a combination of re-shuffled \DBeta, plus additional data. The goal of having different types of datasets is to study the generalizability of \GS in detecting code coming from heterogeneous sources.


\begin{table}[h!]
	\vspace{-.1cm}
	\centering
	\small
	\footnotesize
	\caption{Number of collected code snippets.} 
	\vspace{-.2cm}
	\begin{tabular}{|p{1.5cm}|p{2.8cm}|p{2.8cm}|}    \hline  
	 \textbf{Source}	 & \textbf{Unpaired snippets $\mathcal{U}$ (Section \ref{sec:IndependentSnippets})}  & \textbf{Paired snippets $\mathcal{P}$ (Section \ref{sec:PairwiseSnippets})} \\ \hline 
		\CG & 137  & \numGPTSnippets \\ \hline 
		Humans  & 137  & \numHumanSnippets \\ \hline \hline 
		 \multicolumn{3}{|c|}{\textbf{Dataset}} \\ \hline
		\DAlfa = $\mathcal{U} \cup \mathcal{P}$  & \multicolumn{2}{c|}{1,484} \\ \hline
		\DBeta $\equiv$ $\mathcal{P}$  &  & 1,210 \\ \hline 
	\end{tabular}	
	\label{tab:Datasets}
	\vspace{-.1cm}
\end{table}





%

\subsection{Comparison with the Baselines}
The comparison with the baselines was performed using the paired snippets
$\mathcal{P}$ (see Section~\ref{sec:PairwiseSnippets}) for which each human-written snippet has a counterpart generated by \CG.



\GZ and \OT are different with respect to the length of the input data they can handle. \OT accepts only text with more than 1,000 characters, and \GZ can handle shorter text with at least 250 characters. For this reason, we had to create two separate lists of queries. In particular, for comparing \GZ with \GS, 50 snippets of small size were chosen for each of the two categories ``\emph{Human},'' and ``\emph{\CG}.'' For comparing \OT with \GS, there were 50 snippets of more than 1,000 characters for each of the two categories ``\emph{Human},'' and ``\emph{\CG}.'' 


\subsection{Evaluation Settings and Metrics} \label{sec:SettingsAndMetrics}
We split the data using the 80:10:10 ratio, \ie 80\%, 10\%, and 10\% of the data are used for training, validation, and testing, respectively. For each testing snippet, before being fed as input to the prediction engine, its real category, \ie either ``\emph{Human}''  or ``\emph{\CG}'' is removed to use as ground-truth data. For every testing snippet, we evaluated it by comparing its actual category with the predicted one returned by \GS, and computed the number of True positives (TP), False positives (FP), False negatives (FN), and True negatives (TN)~\cite{Dalianis2018}. The final performance is evaluated using Accuracy, Precision, Recall, and F$_1$-score, defined as follows. 



\subsubsection{Accuracy} It measures the ratio of correctly classified snippets to the total number of snippets for all the considered categories, computed as follows.
\vspace{-.1cm}
\begin{equation} \label{eqn:Accuracy} \nonumber%
	Accuracy = \frac{\left | TP+TN \right |}{\left | TP+TN+FP+FN \right |}
\end{equation}

\subsubsection{Precision, Recall, and F$_1$-score} 
Given a category, Precision measures the fraction of correctly classified items to the total number of items; 
Recall is the ratio of actual positive cases that are correctly classified; F$_1$-score (or F$_1$) is a harmonic combination of the two aforementioned metrics. 
\vspace{.1cm}

\begin{minipage}{0.3\linewidth}
	\begin{equation} \nonumber
		P = \frac{\left | TP \right |}{\left | TP+FP \right |};
	\end{equation}
\end{minipage} 
\begin{minipage}{0.3\linewidth}
	\begin{equation} \nonumber
		R = \frac{\left | TP \right |}{\left | TP+FN \right |}; 
	\end{equation}
\end{minipage}
\begin{minipage}{0.3\linewidth}
	\begin{equation} \nonumber
		F_1 = \frac{2\times P\times R}{P+R} 
	\end{equation}
\end{minipage}

\vspace{.2cm}
In the evaluation, we also make use of \emph{macro average}, and \emph{weighted average} score of these metrics. The former is the arithmetic mean of all the scores for the two categories, while the latter weighs the varying degree of importance of the categories in a dataset.

In RQ$_3$, to compare \GS with \GZ and \OT, we use McNemar's test \cite{mcnemar}, which is a proportion test for paired samples. As we perform multiple comparisons, $p$-values are adjusted using Holm's correction \cite{Holm1979a}. The McNemar's test is complemented by the Odds Ratio (OR) effect size measure.




\subsection{Configurations}\label{sec:configuration}

\begin{figure*}[t!]
	\centering
	\begin{tabular}{ccc}	
		\subfigure[C$_1$]{\label{fig:ConfigurationC1}
			\includegraphics[width=0.30\linewidth]{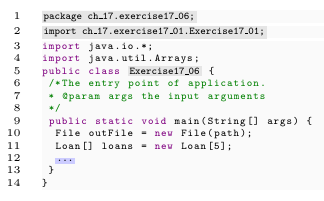}} &		
		\subfigure[C$_2$]{\label{fig:ConfigurationC2}
			\includegraphics[width=0.30\linewidth]{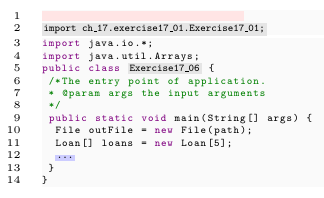}} &		
		\subfigure[C$_3$]{\label{fig:ConfigurationC3}
			\includegraphics[width=0.30\linewidth]{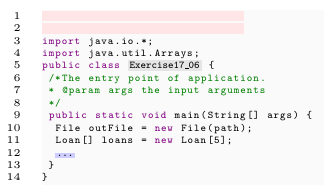}} \\
		
		\subfigure[C$_4$]{\label{fig:ConfigurationC4}
			\includegraphics[width=0.30\linewidth]{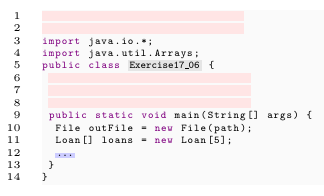}} &				
		\subfigure[C$_5$]{\label{fig:ConfigurationC5}
			\includegraphics[width=0.30\linewidth]{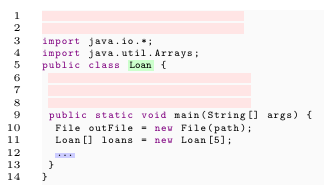}} &
		\subfigure[C$_6$]{\label{fig:ConfigurationC6}
			\includegraphics[width=0.30\linewidth]{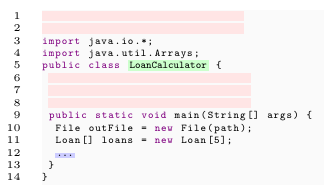}}			
	\end{tabular} 
	\vspace{-.2cm}
	\caption{Different configurations for human-written code.} 	
	\label{fig:CodeExampleAndModifications}
	\vspace{-.4cm}
\end{figure*}

%



Table~\ref{tab:Configurations} shows the \numConf experimental configurations, which are indeed not exhaustive, as we cannot consider all possible combinations of artifacts. Thus, we pay attention only to those most representative and realistic, as explained below. We use the check mark symbol \faCheck~to indicate that the corresponding feature is kept; and a uncheck mark symbol \untick~to signal the opposite, \ie removing the feature; the hand-written symbol \edited~represents a modification in the feature, where \edited{\footnotesize G} means the original name is replaced by that coming from the corresponding \CG snippet, and \edited{\footnotesize H} signals that such a name is replaced by humans. 

\begin{table}[h!]
	\centering
	\scriptsize	
	\vspace{-.1cm}
	\caption{Experimental configurations.}
	\begin{tabular}{|p{2.0cm}| l | l | l | l |l |l |l|l|}	\hline
		 \textbf{Artifact}  & \multicolumn{8}{c|}{\textbf{Configurations}} \\ \hline
		 & \textbf{C$_1$} & \textbf{C$_2$} & \textbf{C$_3$}& \textbf{C$_4$} & \textbf{C$_5$} & \textbf{C$_6$}  &C$_7$ & C$_8$ \\ \hline
		Package definition  & \faCheck & \untick & \untick & \untick & \untick  & \untick & \untick& \untick\\ \hline
		Self-made class name &  \faCheck & \faCheck & \faCheck  & \faCheck & \edited \scriptsize G& \edited \scriptsize H & \edited \scriptsize H & \edited \scriptsize H \\ \hline
		Imports to self-made packages &  \faCheck & \faCheck & \untick & \untick & \untick & \untick & \untick & \untick \\ \hline
		Code comments  & \faCheck & \faCheck & \faCheck & \untick & \untick  & \untick & \untick & \untick \\ \hline
		All imports & \faCheck & \faCheck & \faCheck & \faCheck & \faCheck & \faCheck & \untick & \untick\\ \hline

		\texttt{\textbackslash t} and \texttt{\textbackslash n} & \faCheck & \faCheck & \faCheck & \faCheck & \faCheck & \faCheck & \faCheck & \untick\\ \hline
		
	\end{tabular}
	\vspace{-.2cm}
	\label{tab:Configurations}
\end{table}


By default, \CG never generates a package name (we noticed this after several attempts of interacting with the platform), thus with package definition, we only consider snippets written by humans. 
Fig.~\ref{fig:CodeExampleAndModifications} illustrates different snippets corresponding to the considered configurations, explained as follows:

\begin{itemize}
	\item \textbf{C$_1$}: We keep the code by \CG and human unchanged, and run the experiments with the code as it is. An example of such code is shown in Fig.~\ref{fig:ConfigurationC1}. 
	\item \textbf{C$_2$}: In this configuration, the package definition from the human-written code is removed. As shown in Fig.~\ref{fig:ConfigurationC2}, compared to Fig.~\ref{fig:ConfigurationC1}, \api{package\; ch\_17.exercise17\_06} is no longer seen. 
	\item \textbf{C$_3$}: From \textbf{C$_2$}, imports to project packages are dropped. The code in Fig.~\ref{fig:ConfigurationC3} is similar to that in Fig.~\ref{fig:ConfigurationC2}, however the import directive \api{import\; ch\_17.exercise17\_01.Exercise17\_01} is removed.
	\item \textbf{C$_4$}: From C$_3$, comments embedded in the code by of humans and \CG are deleted. Fig.~\ref{fig:ConfigurationC3} shows the snippet written by humans but without code comments.
	\item \textbf{C$_5$}: We conjecture that strings related to the hierarchical names, \eg \api{Exercise17\_06}, might be a discriminant feature, creating a bias in the prediction performance. Thus, from the human-written code in \textbf{C$_4$}, we replace the class name with that of the corresponding snippet written by \CG (Fig.~\ref{fig:ConfigurationC5}). 
	\item \textbf{C$_6$}: We attempt to make the code more human-like by giving a name that reflects well the task. From the human-written code in \textbf{C$_4$}, the co-authors of this paper read the task assignment, and renamed the class with a name that reflects better the task (Fig.~\ref{fig:ConfigurationC6}). 
	\item \textbf{C$_7$}: From the human-written code in \textbf{C$_6$}, and the \CG-generated code in \textbf{C$_3$}, we removed all the imports statements. 
	\item \textbf{C$_8$}: From the human-written and the \CG-generated code in \textbf{C$_7$}, we removed all the formatting characters, including $\backslash$\texttt{t} and $\backslash$\texttt{n}.  Due to space limits, we do not display a figure to illustrate the code examples for \textbf{C$_7$} and \textbf{C$_8$}. Further examples are in our online appendix \cite{replication}.
\end{itemize}

In the experiments, we executed \GS on the datasets along the aforementioned configurations. The obtained results are reported and analyzed in the next section.



%% file: src/Results.tex

In the following we report the results of the study addressing the research questions formulated in Section \ref{sec:Evaluation}

\subsection{\rqfirst}
\label{sec:RQ1}

\input{src/RQ1}

\subsection{\rqsecond}
\label{sec:RQ2}
\input{src/RQ2}

\subsection{\rqthird}
\label{sec:RQ3}
\input{src/RQ3}

%% file: src/RQ1.tex

In this research question, we study the effect of the training data on the accuracy of \GS. To this end, we consider two use cases: \emph{(i)} \textit{Testing and training data come from independent sources}; and \emph{(ii)} \textit{Testing and training data come from same sources}. 
Note that in the following the term ``independent'' means that snippets come from completely different datasets. It is always the case that there is no duplicate between the training and test set.

\subsubsection{Testing and training data come from independent sources}

We first investigate how \GS performs when it is tested from a dataset coming from a completely different source/domain than the training set.
To this aim, we use the unpaired snippets $\mathcal{U}$ (see Section~\ref{sec:IndependentSnippets}) to test the system, which has already been trained with the paired snippets, \ie $\mathcal{P}$ (see Section~\ref{sec:PairwiseSnippets}).
%
Our goal is to replicate a realistic scenario where \GS will make predictions on snippets that were not used during training.

Table~\ref{tab:IndependentData} reports the evaluation metrics for this experiment, considering the dataset without any preprocessing, \ie C$_1$. The \textbf{Support (\#)} column indicates the number of testing items for each category.
Overall,  \GS obtains a low prediction performance for both categories. While it achieves 1.00 as Recall for code written by \CG, it yields 0.57 as Precision, resulting in 0.73 as F$_1$ score for the set of 120 testing instances. When detecting code written by humans, \GS also achieves a low Recall, \ie 0.25, thus decreasing the corresponding F$_1$ score to 0.40.



\begin{table}[h!]
	\vspace{-.1cm}
	\centering
	\small
	\footnotesize
	\caption{$\mathcal{P}$ for training, and $\mathcal{U}$ for testing.} 
	\vspace{-.1cm}
	\begin{tabular}{|p{1.5cm}|p{1.1cm}|p{1.1cm}|p{1.1cm}|p{1.4cm}|}    \hline  
		& \textbf{Precision} & \textbf{Recall} & \textbf{F$_1$ score} & \textbf{Support (\#)} \\ \hline 
		\CG & 0.57 & 1.00 & 0.73 & 120 \\ \hline 
		Humans & 1.00 & 0.25 & 0.40 & 120 \\ \hline \hline
		
		accuracy &  &  & 0.62 & 240 \\ \hline 
		macro avg &  0.79 & 0.62 & 0.56 & 240 \\ \hline 
		weighted avg &  0.79 & 0.62 & 0.56 & 240 \\ \hline 
	\end{tabular}	
	\label{tab:IndependentData}
\end{table}

In summary, the empirical evidence indicates that in the presence of completely different data sources between training and test sets, the prediction becomes challenging and results in mediocre performance.

\subsubsection{Testing and training data come from same sources}

For this experiment, we use 
\DAlfa, where snippets in $\mathcal{P}$ and $\mathcal{U}$ are mixed and split again to distribute the snippets coming from different sources into balanced parts. 
The execution of \GS on \DAlfa produced the results reported in Tables \ref{tab:Precision_DB}, \ref{tab:Recall_DB}, and \ref{tab:F1_DB} (with the best results being printed in bold), analyzed as follows.

%


	%

\begin{table}[h!]
	\centering
	\small
	\footnotesize
	\caption{\DAlfa: Precision.}  
		\label{tab:Precision_DB}
	\vspace{-.1cm}
	
	\begin{tabular}{|p{1.5cm}|p{0.35cm}|p{0.35cm}|p{0.35cm}|p{0.35cm}|p{0.35cm}|p{0.35cm}|p{0.35cm}|p{0.35cm}|p{0.35cm}|}  \hline  
		& \textbf{C$_1$} & \textbf{C$_2$} & \textbf{C$_3$} & \textbf{C$_4$} & \textbf{C$_5$} & \textbf{C$_6$} & \textbf{C$_7$} & \textbf{C$_8$} & \textbf{\#} \\ \hline 
		\CG & 0.90 & 0.88 & 0.91 & 0.92 & 0.93 & 0.91 & 0.86 & 0.84 & 148 \\ \hline 
		Humans & 0.98 & \textbf{1.00} & \textbf{1.00} & \textbf{1.00} & 0.99 & \textbf{1.00} & \textbf{1.00} & 0.98 & 147 \\ \hline \hline
		
		macro avg & 0.94 & 0.94 & 0.96 & 0.96 & 0.96 & 0.95 & 0.93 & 0.91 & 295 \\ \hline 
		weighted avg & 0.94 & 0.94 & 0.96 & 0.96 & 0.96 & 0.95 & 0.93 & 0.91 & 295 \\ \hline 
	\end{tabular}	
	\vspace{-.1cm}
\end{table}

As shown in Table~\ref{tab:Precision_DB}, 
\GS is more precise in identifying human-written code than \CG-generated code. For five out of \numConf configurations, \GS achieves 1.00 as Precision, and by the remaining three configurations, the corresponding values are 0.98, 0.99, and 0.98. Concerning macro and weighted average Precision scores, for C$_8$ the tool yields the lowest performance, \ie 0.91 as the overall macro average and weighted average Precision.

Table~\ref{tab:Recall_DB} reports the Recall scores for all the considered configurations. Over 148 \CG-generated snippets, \GS yields a Recall of 1.00 for five out of \numConf configurations. At the same time, the corresponding scores obtained on the code written by humans are a bit lower. In particular, for four configurations, the Recall values for this category are below 0.90.

\begin{table}[h!]
	\centering
	\small
	\footnotesize
		\vspace{-.1cm}
	\caption{\DAlfa: Recall.}  
		\label{tab:Recall_DB}
	\begin{tabular}{|p{1.5cm}|p{0.35cm}|p{0.35cm}|p{0.35cm}|p{0.35cm}|p{0.35cm}|p{0.35cm}|p{0.35cm}|p{0.35cm}|p{0.35cm}|}  \hline  
		& \textbf{C$_1$} & \textbf{C$_2$} & \textbf{C$_3$} & \textbf{C$_4$} & \textbf{C$_5$} & \textbf{C$_6$} & \textbf{C$_7$} & \textbf{C$_8$} & \textbf{\#} \\ \hline 
		\CG & 0.99 & \textbf{1.00} & \textbf{1.00} & \textbf{1.00} & 0.99 & \textbf{1.00} & \textbf{1.00} & 0.98 & 148 \\ \hline 
		Humans & 0.89 & 0.86 & 0.90 & 0.91 & 0.93 & 0.90 & 0.83 & 0.82 & 147 \\ \hline \hline
		
		macro avg & 0.94 & 0.93 & 0.95 & 0.96 & 0.96 & 0.95 & 0.91 & 0.90 & 295 \\ \hline 
		weighted avg & 0.94 & 0.93 & 0.95 & 0.96 & 0.96 & 0.95 & 0.92 & 0.90 & 295 \\ \hline 
	\end{tabular}	
	\vspace{-.1cm}
\end{table}

Table~\ref{tab:F1_DB} reports the accuracy and F$_1$-scores obtained for all the configurations. Overall, the F$_1$-scores are greater than or equal to 0.90. Among the configurations, \GS gets the best accuracy, \ie 0.96, by C$_4$ and C$_5$. This implies that once all the package definitions and code comments have been removed from the original snippets, \GS improves its performances. One possible interpretation of this phenomenon is that, on the one hand, package definitions just include recurring items present on both human-written and \CG-generated code. On the other hand, comments (and natural language elements in general) may not contain features that \GS can leverage to successfully perform a classification.

\begin{table}[h!]
	\centering
	\small
	\footnotesize
	\caption{\DAlfa: Accuracy and F$_1$ score.}  
		\label{tab:F1_DB} 
	\vspace{-.1cm}
	\begin{tabular}{|p{1.5cm}|p{0.35cm}|p{0.35cm}|p{0.35cm}|p{0.35cm}|p{0.35cm}|p{0.35cm}|p{0.35cm}|p{0.35cm}|p{0.35cm}|}  \hline  
		& \textbf{C$_1$} & \textbf{C$_2$} & \textbf{C$_3$} & \textbf{C$_4$} & \textbf{C$_5$} & \textbf{C$_6$} & \textbf{C$_7$} & \textbf{C$_8$} & \textbf{\#} \\ \hline 
		\CG & 0.94 & 0.94 & 0.95 & 0.96 & 0.96 & 0.95 & 0.92 & 0.91 & 148 \\ \hline 
		Humans & 0.94 & 0.93 & 0.95 & 0.95 & 0.95 & 0.95 & 0.91 & 0.89 & 147 \\ \hline \hline
		
		accuracy & 0.94 & 0.93 & 0.95 & \textbf{0.96} & \textbf{0.96} & 0.95 & 0.92 & 0.90 & 295 \\ \hline 
		macro avg & 0.94 &  0.93 & 0.95 & \textbf{0.96} & \textbf{0.96} & 0.95  & 0.91 & 0.90 & 295 \\ \hline 
		weighted avg & 0.94 & 0.93 & 0.95 & \textbf{0.96} & \textbf{0.96} & 0.95 & 0.91 & 0.90 & 295 \\ \hline 
	\end{tabular}	
	\vspace{-.2cm}
\end{table}

To sum up, on the one hand, \GS does not perform well when tested on code belonging to a completely different domain dataset. But on the other hand, its performance considerably improves when the common patterns--those that may occur in data curated from the same domains--have been learned during the training. 


\vspace{.1cm}
\begin{shadedbox}
	\small{\textbf{Answer to RQ$_1$.} \GS cannot recognize well data originating from completely extraneous sources than the one it has been trained with. At the same time, its performance metrics are all at least 90\% when it is trained with data from a source seen before.}
\end{shadedbox}
\vspace{1mm}

%% file: src/RQ2.tex






After having investigated how \GS performs on completely different training and test sets, as well as on related ones, we experiment with the most favorable conditions, \ie the use of a paired dataset (\DBeta), for which each human-written snippet is associated with a corresponding version generated by \CG.

For this dataset, we achieve an almost perfectly consistent outcome, \ie by six out of eight configurations, \ie C$_1$, C$_2$, C$_3$, C$_5$, C$_6$, and C$_7$, the accuracy is 1.0, so are Precision, Recall, and F$_1$-score. For the sake of clarity, we report the results for these configurations using a single table, \ie Table~\ref{tab:C1_C7}. 

\begin{table}[h!]
	\vspace{-.1cm}
	\centering
	\small
	\footnotesize
	\caption{\DBeta: C$_1$, C$_2$, C$_3$, C$_5$, C$_6$, and C$_7$.}   %
	\begin{tabular}{|p{1.5cm}|p{1.1cm}|p{1.1cm}|p{1.1cm}|p{1.4cm}|}    \hline  
		& \textbf{Precision} & \textbf{Recall} & \textbf{F$_1$ score} & \textbf{\#} \\ \hline 
		\CG & 1.00 & 1.00 & 1.00 & 120 \\ \hline 
		Humans & 1.00 & 1.00 & 1.00 & 120 \\ \hline \hline	
		accuracy &  &  & 1.00 & 240 \\ \hline 
		macro avg & 1.00 & 1.00 & 1.00 & 240 \\ \hline 
		weighted avg & 1.00 & 1.00 & 1.00 & 240 \\ \hline 
	\end{tabular}	
	\label{tab:C1_C7}
	\vspace{-.1cm}
\end{table}

As it can be noticed, by C$_4$ and C$_8$, \GS exhibits a slightly lower performance compared to that of the other configurations as shown in Table~\ref{tab:C4_and_C8}. In particular, the tool always gets 0.99 as macro average and weighted average for Precision, Recall, and F$_1$ score. In these configurations, \GS is better at detecting code by humans, compared to code by \CG, \ie the obtained Precision, Recall, and F$_1$ score are 1.00, 0.99, and 0.99 for human code. The corresponding scores for code generated by \CG are 0.99, 0.98, and 0.99. Still, it is clear that \GS is able to detect well code written by \CG and humans in the two aforementioned configurations. 


\begin{table}[h!]
	\vspace{-.1cm}
	\centering
	\small
	\footnotesize
	\caption{\DBeta: C$_4$ and C$_8$.}  
	\begin{tabular}{|p{1.5cm}|p{1.1cm}|p{1.1cm}|p{1.1cm}|p{1.4cm}|}    \hline  
		& \textbf{Precision} & \textbf{Recall} & \textbf{F$_1$ score} & \textbf{\#} \\ \hline 
		\CG & 0.99 & 0.98 & 0.99 & 120 \\ \hline 
		Humans & 1.00 & 0.99 & 0.99 & 120 \\ \hline \hline
		
		accuracy &  &  & 0.99 & 240 \\ \hline 
		macro avg & 0.99 & 0.99 & 0.99 & 240 \\ \hline 
		weighted avg & 0.99 & 0.99 & 0.99 & 240 \\ \hline 
	\end{tabular}	
	\label{tab:C4_and_C8}
	\vspace{-.1cm}
\end{table}

From the results of Table~\ref{tab:C1_C7} and Table~\ref{tab:C4_and_C8}, we can conclude that, for the paired dataset, \GS can properly tell apart code written by humans and AI. 
While the results of this scenario seem obvious, they acquired knowledge that can be leveraged to properly train \GS for applications in which it is expected to obtain several similar code snippets, \eg to detect plagiarism in courses' assignments.

%
%
%

\vspace{.1cm}
\begin{shadedbox}
	\small{\textbf{Answer to RQ$_2$.} On the paired dataset, \GS obtains a perfect prediction by the majority of the experimental settings. This indicates that when being trained with pairwise code, \GS works well as a detector for code written by \CG.} 
\end{shadedbox}
\vspace{1mm}


%% file: src/RQ3.tex
 
 In the following we report the comparison of \GS with \GZ and \OT. For \GS, we selected the two most representative configurations, \ie C$_1$ and C$_8$, as they correspond to the cases when \GS performs the best and the worst, respectively (see Table~\ref{tab:C1_C7} and Table~\ref{tab:C4_and_C8}). 




\subsubsection{Comparison with \GZ}


The classification results returned by \GZ are shown in Table~\ref{tab:GPTZeroResults}. The first column reports the answer text returned by \GZ, the second column shows a binary classification given by us to allow for a comparison with \GS, and the third column shows the number of queries. Most of the queries, \ie 81 + 7 snippets, are classified as written by humans, and only 6 + 6 snippets are predicted as generated by \CG. 




\begin{table}[t]
	\caption{Classification results by \GZ.}
	\label{tab:GPTZeroResults}
	\footnotesize
	\begin{tabular}{|p{5.8cm}|l|l|} \hline
		\textbf{Answer}                  & \textbf{Final class}      & \textbf{\#} \\ \hline
		Your text is likely to be written entirely by a human                          &   Humans      & 81  	 \\ \hline
		Your text is most likely human written but there are some sentences with low perplexities & Humans  &  7       \\ \hline
		Your text is likely to be written entirely by AI                                          & \CG  &  6       \\ \hline
		Your text may include parts written by AI                                                 & \CG  &  6       \\ \hline		                                              
	\end{tabular}
	\vspace{-4mm}
\end{table}

By matching with the ground-truth data, we obtained the following classification results: 64 out of 100 snippets are correctly predicted by \GZ, corresponding to an Accuracy of 0.64.  The outcome obtained by \GS is as follows: among 100 snippets, 99 are correctly classified, \ie Accuracy = 0.99. McNemar's test indicates a statistically significant difference ($p$-value$<2e-09$), with an OR=42 in favor of \GS.
\emph{This means that \GS significantly outperforms \GZ in recognizing code generated by \CG}.

\subsubsection{Comparison with \OT}

Table~\ref{tab:OpenAIResults} depicts the classification results obtained by running with \OT. Similar to Table~\ref{tab:GPTZeroResults}, the first column depicts the answer text given by \OT, the second column is the binary final classification given by us to compare with \GS, and the third column represents the number of queries. Among them, 35 + 3 snippets are marked as written by humans, and the rest, \ie 20 + 42 snippets are predicted by \OT as generated by \CG. 

\begin{table}[h!]
	\vspace{-1mm}
	\caption{Classification results by \OT.}
	\label{tab:OpenAIResults}
	\footnotesize
	\begin{tabular}{|p{5.8cm}|l|l|} \hline
		\textbf{Answer}                     & \textbf{Final class}  & \textbf{\#} \\ \hline
		The classifier considers the text to be unclear if it is AI-generated &   Humans   & 35  \\ \hline
		The classifier considers the text to be unlikely AI-generated         &   Humans   & 3   \\ \hline
		The classifier considers the text to be likely AI-generated           & \CG  & 20  \\ \hline
		The classifier considers the text to be possibly AI-generated         & \CG  & 42  \\ \hline
\end{tabular}
\end{table}

Comparing with the real category of the query snippets, we see that 61 out of 100 are correctly predicted by \OT, resulting in an Accuracy of 0.61. The classification by \GS is almost perfect: 99 out of 100 snippets are correctly recognized, corresponding to an Accuracy of 0.99.  McNemar's test indicates a statistically significant difference ($p$-value$<2e-09$), with an OR=38 in favor of \GS.

Altogether, we can conclude that \emph{the prediction by \GS is significantly better than the one provided by \OT}. Noteworthy, \OT has a disclaimer saying that: ``\emph{The classifier isn't always accurate; it can mislabel both AI-generated and human-written text. AI-generated text can be edited easily to evade the classifier.}'' 



\begin{shadedbox}
	\small{\textbf{Answer to RQ$_3$.} \GS considerably outperforms both \GZ and \OT in the ability to detect if a code snippet is generated by \CG.}
\end{shadedbox}
\vspace{1mm}

%% file: src/Discussion.tex
We discuss possible implications from the experimental results, 
and highlight the threats to the validity of our findings.

\subsection{Implications}

The first result emerging from our study is that training performed on completely different data may lead to sub-optimal results. This is analogous to what happens to other kinds of predictors in software engineering, for example defect prediction models, for which a cross-project prediction performs well only when the training and test feature closely-related (\eg in terms of metrics) code elements \cite{MenziesBCMLSTZ13}.

\begin{shadedbox}
	\small{\textbf{Finding 1.} To effectively recognize AI-generated code, an classifier such as \GS needs to be trained on source code being relevant to the context under consideration.
	}
\end{shadedbox}

By experimenting \GS with different configurations, we noticed how certain preprocessing, such as removing comments or package names actually helps to improve the performance of the classification. While future work is needed to perform a feature importance analysis, the obtained results suggest that, in general, it is useful to apply aggressive preprocessing to make the learning model better generalizable and perform better.

\begin{shadedbox}
	\small{\textbf{Finding 2.} ML-model to recognize AI-generated code should properly preprocess the input source code in order to achieve good results and be generalizable.}
\end{shadedbox}

We found out that \GS works almost perfectly for paired code elements, even in the presence of an aggressive preprocessing. While such a usage scenario is not common in development activities, it may occur an educational environment. In such a case, a teacher may have different instances of the same source code, \eg by the book, produced by themselves, or by different students, along with solutions generated by an AI. 

\begin{shadedbox}
	\small{\textbf{Finding 3.} Whenever possible, \eg in educational scenarios, one can train an ML model with paired snippets to achieve almost perfect predictions.}
\end{shadedbox}






%
%
%

\subsection{Threats to validity}

\emph{Construct validity} concerns the relationship between theory and observation. In principle, the dataset labeling is correct ``by construction'' as we know a priori its origin. For code hosted on GitHub,  \CG may have seen it already. However, this does not necessarily bias our study, as \CG \cite{chatgpt} generates code rather than retrieving snippets relevant to a query, introducing its peculiar (and recognizable) elements, \eg imports, formatting, or other code-style features.
Along this line, the different preprocessing configurations C$_1$--C$_8$ simulate the different ways a generative model could make its code ``looking different'' than the human code it may have been trained with. Another threat could be that for RQ$_3$ we mapped the four categories provided by \GZ and \OT to binary categories. We used the ``likely/possibly'' written by \CG as \CG category.

\emph {Internal validity} threats concern factors, internal to our study, that can influence the results. We ran the queries directly with \GZ and \OT using their Web interface. As this makes the process time-consuming, we ran each query once, yet it is possible that multiple runs would produce different results. Concerning the set of hyperparameters to train \GS, we tried with different combinations of batch size, warm-up steps, or weight decay, so as to rule out possible internal threats.

\GS has been implemented using \CB \cite{feng-etal-2020-codebert}, even though other alternatives--possibly better code-pre-trained models such as CodeT5 \cite{wang2021codet5}--do exist. That being said, \emph{(i)} \CB already reached very good performance, outperforming existing classifiers such as \GZ and \OT; \emph{(ii)} the goal of our work was not to develop the best AI-code detector possible, but rather, to show how a specific fine-tuning of a code-pretrained model works better than just leveraging out-of-the-box classifiers. In this respect, future work may further improve \GS with better models.


\emph{External validity} threats concern the generalizability of our findings. The findings of this paper may be valid only for the given datasets. We diversified the data by collecting it from different sources, attempting to simulate real-world scenarios. Also, in the evaluation, we experimented with a method definition, not a whole software project. By properly training it, \GS can be adapted to work at the project level. Finally, to avoid adding a further variable to the study, we focused on Java, yet further studies with other languages are highly desirable.

%% file: src/RelatedWork.tex

This section reviews work related to identifying automatically-generated artifacts, and applications of \CG and \CB to software engineering problems.

\subsection{Distinguishing human-written and automated artifacts}

Cassee \etal \cite{9609135} compared three different models to detect whether a GitHub account is human or bot-based using the repositories' comments.
Their evaluation show that combining the textual data with account metadata information leads to better performance, although detecting mixed accounts is still challenging, \ie the best configuration identifies bot activity in only 10\% of mixed accounts.
On the same line, BIMAN \cite{10.1145/3379597.3387478} is a hybrid approach that identifies bot accounts on GitHub by relying on three different features, \ie the names of the account, the commit messages, and associations between commits and projects. 
The evaluation showed that BIMAN succeeds in recognizing bots with good accuracy.

Paltenghi and Pradel \cite{9678712} compared the neural network attention mechanism, and the human one. By focusing on code summarization tasks, the authors collected 1,508 human records and extracted 250 labeled methods by 91 participants. Their experiments showed that the neural attention mechanism  \textit{(i)} struggles in recognizing the longer methods, and \textit{(ii)} underestimates the value of strings in the code. 

Morales \etal \cite{morales_repor_2020} performed an empirical study investigating whether automated refactorings can be as effective as human ones. To this end, the authors involved 80 developers in classifying 20 refactoring tasks, including human-written code and refactorings generated by RePRO--a state-of-the-art tool. The study results showed that developers cannot identify automated refactoring by five different anti-patterns.

Glzadeh \etal \cite{9474384} compared several classifiers to identify bots from issue and pull request comments. To this end, the input data collected from a pre-labeled dataset is encoded by combining bag of words and TF-IDF indexing. Their results show that Multinomial Naive Bayesian outperforms the others in terms of precision, recall, and F1 score.     

Particularly relevant to our work are alternative approaches to detect text generated by \CG, \ie, \GZ and \OT.
\GZ \cite{gptzero} which is an academic tool specifically conceived to detect text generated by \CG. Given a textual content between 250 and 5,000 characters, the underpinning model can categorize can distinguish if the snippet belongs to \CG, human, or mixed implementation. An alternative to \GZ is \OT \cite{ot}, a tool developed by OpenAI. The results of the conducted evaluation show that our approach outperforms both \GZ and \OT for the queries we considered.


\subsection{Applications of \CG in Software Engineering}

Software Engineering studies have recently been placing more emphasis on \CG. Ahmad \etal \cite{ahmad2023humanbot} tested the effectiveness of \CG in aiding a software architect. The study leveraged \CG for analyzing, synthesizing, and evaluating a services-oriented software application's architecture.
The effectiveness of \CG as coding assistance has been explored by Chauvet \etal~\cite{avila_chauvet_chatgpt_2023}, using \CG to get help with developing in HTML, CSS, and JavaScript. 

Sobania \etal~\cite{sobania2023analysis} studied the performance of \CG in fixing bugs. They compared \CG with state-of-the-art tools. The results indicated that \CG performs bug-fixing tasks successfully in most cases (31 of 40 bugs tested).

Cao~\etal \cite{cao2023study} explored the feasibility of employing \CG for deep learning program repair with fault detection and fault localization. Furthermore, they investigated the impact of prompts on debugging performance and eventually proposed a template to achieve better results.


The effectiveness of \CG as coding assistance has been explored by Chauvet \etal~\cite{avila_chauvet_chatgpt_2023}, which leveraged \CG to get assistance with developing in HTML, CSS and JavaScript.

To the best of our knowledge, \GS is the first attempt to employ a pre-trained model to detect if a code snippet is written by humans or generated by \CG. 

\subsection{Applications of code pre-trained models to Software Engineering Tasks}

The success of pre-trained models NLP 
has led to the development of similar models for programming language understanding and generation. Examples of such models include CodeBERT \cite{feng-etal-2020-codebert}, GraphCodeBERT \cite{GraphCodeBERT_2021}, PLBART \cite{Ahmad_Chakraborty_Ray_Chang_2021}, or CodeT5 \cite{wang2021codet5} \cite{T5_Oliveto_Bavota_2021}.
As \GS is based on \CB, in the following we discuss recent work based on such a model. For a more comprehensive review of deep learning in software engineering, readers can refer to a survey \cite{10.1145/3485275}.


Wang \etal \cite{wang_bridging_2022} improved \CB models for discriminative code tasks by combining data augmentation and curriculum-learning strategy. 
Their results confirm that the
preprocessing pipeline proposed by Wang \etal  increases \CB's performance in three code-related tasks, \ie algorithm classification, code clone detection, and code search. 

An empirical evaluation of several pre-trained models for code diagnostic tasks, called probes, has been conducted by Karmakar and Robbes \cite{karmakar_what_2022}. To enable the comparison, Karmakar and Robbes  reuse a labeled dataset of compilable Java projects categorized in terms of four different probing tasks, \ie syntactic, surface, semantic, and structure. The outcome of their study shows that \CB is effective in classifying code snippets using semantic information.    



The CCBERT approach \cite{ZHANG2022106922} combines Copy mechanism with \CB to support the generation of enhanced Stack Overflow questions. After a preparatory phase in which bi-modal information is encoded, \CB generates the questions by using a copy attention layer to improve the results that outperform notable baselines.

Gu \etal proposed AdaMO \cite{gu_assemble_2022}, an automatic code summarization tool based on 
GPT-2. AdaMO uses adaptive learning strategies, \ie continuous pre-training and intermediate fine-tuning, to increase the overall performance and Gaussian noise strategy to capture contextual information. Compared with state-of-the-art approaches, AdaMO achieves better results in terms of
ROUGE, METEOR, and BLEU scores.     

%% file: src/Conclusion.tex
Since its release, \CG revealed itself as promising to support various software development tasks, and in particular to create software artifacts, including source code that meets natural language specifications. At the same time, it has emerged the need for techniques and tools that can help users distinguish between automatically generated and human-specified content.

This paper presented an empirical study to investigate the extent to which it is possible to automatically detect whether a code snippet is written by \CG or humans, as well as the factors that can influence this ability. To achieve this, we present \GS, a novel approach to detecting source code written by \CG. \GS can distinguish code written by humans from \CG-generated code under different experimental settings. Also, it outperforms two tools that recognize AI-generated text, \ie \GZ and \OT.
While experimenting \GS under various configurations, we have identified how different preprocessing, as well as the characteristics of training and test impact on the \GS prediction accuracy.
 
Moving forward, we aim to assess \GS using data from diverse sources, and expand the approach to manage additional SE artifacts such as documentation and bug reports. Last but not least, we anticipate that the application of other pre-trained models for code may further improve the prediction performance of \GS.

%% file: main.bbl
\begin{thebibliography}{10}
\providecommand{\url}[1]{#1}
\csname url@samestyle\endcsname
\providecommand{\newblock}{\relax}
\providecommand{\bibinfo}[2]{#2}
\providecommand{\BIBentrySTDinterwordspacing}{\spaceskip=0pt\relax}
\providecommand{\BIBentryALTinterwordstretchfactor}{4}
\providecommand{\BIBentryALTinterwordspacing}{\spaceskip=\fontdimen2\font plus
\BIBentryALTinterwordstretchfactor\fontdimen3\font minus
  \fontdimen4\font\relax}
\providecommand{\BIBforeignlanguage}[2]{{%
\expandafter\ifx\csname l@#1\endcsname\relax
\typeout{** WARNING: IEEEtranS.bst: No hyphenation pattern has been}%
\typeout{** loaded for the language `#1'. Using the pattern for}%
\typeout{** the default language instead.}%
\else
\language=\csname l@#1\endcsname
\fi
#2}}
\providecommand{\BIBdecl}{\relax}
\BIBdecl

\bibitem{chatgpt}
``{ChatGPT~\url{https://openai.com/blog/chatgpt}}.''

\bibitem{gptzero}
``{GPTZero~\url{https://gptzero.me/}}.''

\bibitem{ot}
``{OpenAI Text
  Classifier~\url{https://platform.openai.com/ai-text-classifier}}.''

\bibitem{replication}
``{Replication Package~\url{https://github.com/MDEGroup/GPTSniffer}}.''

\bibitem{chatgptlaw}
``{The impact of Large Language Models on Law
  Enforcement~\url{https://bit.ly/3V30PJH}}.''

\bibitem{ahmad2023humanbot}
\BIBentryALTinterwordspacing
A.~Ahmad, M.~Waseem, P.~Liang, M.~Fehmideh, M.~S. Aktar, and T.~Mikkonen,
  ``Towards human-bot collaborative software architecting with chatgpt,'' 2023.
  [Online]. Available: \url{https://doi.org/10.48550/arXiv.2302.14600}
\BIBentrySTDinterwordspacing

\bibitem{Ahmad_Chakraborty_Ray_Chang_2021}
\BIBentryALTinterwordspacing
W.~U. Ahmad, S.~Chakraborty, B.~Ray, and K.-W. Chang, ``Unified pre-training
  for program understanding and generation,'' no. arXiv:2103.06333, Apr 2021,
  arXiv:2103.06333 [cs]. [Online]. Available:
  \url{http://arxiv.org/abs/2103.06333}
\BIBentrySTDinterwordspacing

\bibitem{avila_chauvet_chatgpt_2023}
\BIBentryALTinterwordspacing
L.~Avila-Chauvet, D.~Mejía, and C.~O. Acosta~Quiroz,
  ``\BIBforeignlanguage{en}{Chatgpt as a {Support} {Tool} for {Online}
  {Behavioral} {Task} {Programming}},'' Rochester, NY, Jan. 2023. [Online].
  Available: \url{https://papers.ssrn.com/abstract=4329020}
\BIBentrySTDinterwordspacing

\bibitem{bosu2015characteristics}
\BIBentryALTinterwordspacing
A.~Bosu, M.~Greiler, and C.~Bird, ``Characteristics of useful code reviews: An
  empirical study at microsoft,'' in \emph{Proceedings of the International
  Conference on Mining Software Repositories}.\hskip 1em plus 0.5em minus
  0.4em\relax IEEE - Institute of Electrical and Electronics Engineers, May
  2015. [Online]. Available:
  \url{https://www.microsoft.com/en-us/research/publication/characteristics-of-useful-code-reviews-an-empirical-study-at-microsoft/}
\BIBentrySTDinterwordspacing

\bibitem{cao2023study}
\BIBentryALTinterwordspacing
J.~Cao, M.~Li, M.~Wen, and S.~chi Cheung, ``A study on prompt design,
  advantages and limitations of chatgpt for deep learning program repair,''
  2023. [Online]. Available: \url{https://doi.org/10.48550/arXiv.2304.08191}
\BIBentrySTDinterwordspacing

\bibitem{9609135}
\BIBentryALTinterwordspacing
N.~Cassee, C.~Kitsanelis, E.~Constantinou, and A.~Serebrenik, ``Human, bot or
  both? a study on the capabilities of classification models on mixed
  accounts,'' in \emph{2021 IEEE International Conference on Software
  Maintenance and Evolution (ICSME)}, 2021, pp. 654--658. [Online]. Available:
  \url{10.1109/ICSME52107.2021.00075}
\BIBentrySTDinterwordspacing

\bibitem{Dalianis2018}
\BIBentryALTinterwordspacing
H.~Dalianis, \emph{Evaluation Metrics and Evaluation}.\hskip 1em plus 0.5em
  minus 0.4em\relax Cham: Springer International Publishing, 2018, pp. 45--53.
  [Online]. Available: \url{https://doi.org/10.1007/978-3-319-78503-5_6}
\BIBentrySTDinterwordspacing

\bibitem{10.1145/3379597.3387478}
\BIBentryALTinterwordspacing
T.~Dey, S.~Mousavi, E.~Ponce, T.~Fry, B.~Vasilescu, A.~Filippova, and
  A.~Mockus, ``Detecting and characterizing bots that commit code,'' in
  \emph{Proceedings of the 17th International Conference on Mining Software
  Repositories}, ser. MSR '20.\hskip 1em plus 0.5em minus 0.4em\relax New York,
  NY, USA: Association for Computing Machinery, 2020, p. 209–219. [Online].
  Available: \url{https://doi.org/10.1145/3379597.3387478}
\BIBentrySTDinterwordspacing

\bibitem{feng-etal-2020-codebert}
\BIBentryALTinterwordspacing
Z.~Feng, D.~Guo, D.~Tang, N.~Duan, X.~Feng, M.~Gong, L.~Shou, B.~Qin, T.~Liu,
  D.~Jiang, and M.~Zhou, ``{C}ode{BERT}: A pre-trained model for programming
  and natural languages,'' in \emph{Findings of the Association for
  Computational Linguistics: EMNLP 2020}.\hskip 1em plus 0.5em minus
  0.4em\relax Online: Association for Computational Linguistics, Nov. 2020, pp.
  1536--1547. [Online]. Available:
  \url{https://aclanthology.org/2020.findings-emnlp.139}
\BIBentrySTDinterwordspacing

\bibitem{copilot}
{GitHub}, ``{GitHub CoPilot \url{https://github.com/features/copilot}}.''

\bibitem{gist}
------, ``{GitHub Gist \url{https://gist.github.com/discover}}.''

\bibitem{9474384}
\BIBentryALTinterwordspacing
M.~Golzadeh, A.~Decan, E.~Constantinou, and T.~Mens, ``Identifying bot activity
  in github pull request and issue comments,'' in \emph{2021 IEEE/ACM Third
  International Workshop on Bots in Software Engineering (BotSE)}, 2021, pp.
  21--25. [Online]. Available: \url{10.1109/BotSE52550.2021.00012}
\BIBentrySTDinterwordspacing

\bibitem{gu_assemble_2022}
\BIBentryALTinterwordspacing
J.~Gu, P.~Salza, and H.~C. Gall, ``Assemble {Foundation} {Models} for
  {Automatic} {Code} {Summarization},'' in \emph{2022 {IEEE} {International}
  {Conference} on {Software} {Analysis}, {Evolution} and {Reengineering}
  ({SANER})}, Mar. 2022, pp. 935--946, iSSN: 1534-5351. [Online]. Available:
  \url{10.1109/SANER53432.2022.00112}
\BIBentrySTDinterwordspacing

\bibitem{GraphCodeBERT_2021}
\BIBentryALTinterwordspacing
D.~Guo, S.~Ren, S.~Lu, Z.~Feng, D.~Tang, S.~Liu, L.~Zhou, N.~Duan,
  A.~Svyatkovskiy, S.~Fu, M.~Tufano, S.~K. Deng, C.~Clement, D.~Drain,
  N.~Sundaresan, J.~Yin, D.~Jiang, and M.~Zhou, ``Graphcodebert: Pre-training
  code representations with data flow,'' no. arXiv:2009.08366, Sep 2021,
  arXiv:2009.08366 [cs]. [Online]. Available:
  \url{http://arxiv.org/abs/2009.08366}
\BIBentrySTDinterwordspacing

\bibitem{introjavarepo}
\BIBentryALTinterwordspacing
{H. Dulaney}, ``{Solutions to Introduction to Java Programming by Y. Daniel
  Liang. 10th Edition}.'' [Online]. Available:
  \url{https://github.com/HarryDulaney/intro-to-java-programming}
\BIBentrySTDinterwordspacing

\bibitem{Holm1979a}
S.~Holm, ``A simple sequentially rejective multiple test procedure,''
  \emph{Scandinavian journal of statistics}, pp. 65--70, 1979.

\bibitem{DBLP:journals/corr/abs-1909-09436}
\BIBentryALTinterwordspacing
H.~Husain, H.~Wu, T.~Gazit, M.~Allamanis, and M.~Brockschmidt, ``Codesearchnet
  challenge: Evaluating the state of semantic code search,'' \emph{CoRR}, vol.
  abs/1909.09436, 2019. [Online]. Available:
  \url{http://arxiv.org/abs/1909.09436}
\BIBentrySTDinterwordspacing

\bibitem{karmakar_what_2022}
\BIBentryALTinterwordspacing
A.~Karmakar and R.~Robbes, ``What do pre-trained code models know about code?''
  in \emph{Proceedings of the 36th {IEEE}/{ACM} {International} {Conference} on
  {Automated} {Software} {Engineering}}, ser. {ASE} '21.\hskip 1em plus 0.5em
  minus 0.4em\relax Melbourne, Australia: IEEE Press, Jun. 2022, pp.
  1332--1336. [Online]. Available:
  \url{https://dl.acm.org/doi/10.1109/ASE51524.2021.9678927}
\BIBentrySTDinterwordspacing

\bibitem{10.1145/3510003.3510181}
\BIBentryALTinterwordspacing
Z.~Li, G.~Q. Chen, C.~Chen, Y.~Zou, and S.~Xu, ``Ropgen: Towards robust code
  authorship attribution via automatic coding style transformation,'' in
  \emph{Proceedings of the 44th International Conference on Software
  Engineering}, ser. ICSE '22.\hskip 1em plus 0.5em minus 0.4em\relax New York,
  NY, USA: Association for Computing Machinery, 2022, p. 1906–1918. [Online].
  Available: \url{https://doi.org/10.1145/3510003.3510181}
\BIBentrySTDinterwordspacing

\bibitem{liang2003introduction}
Y.~D. Liang, \emph{Introduction to Java programming}.\hskip 1em plus 0.5em
  minus 0.4em\relax Pearson Education India, 2003.

\bibitem{T5_Oliveto_Bavota_2021}
\BIBentryALTinterwordspacing
A.~Mastropaolo, S.~Scalabrino, N.~Cooper, D.~Nader~Palacio, D.~Poshyvanyk,
  R.~Oliveto, and G.~Bavota, ``Studying the usage of text-to-text transfer
  transformer to support code-related tasks,'' in \emph{2021 IEEE/ACM 43rd
  International Conference on Software Engineering (ICSE)}.\hskip 1em plus
  0.5em minus 0.4em\relax Madrid, ES: IEEE, May 2021, pp. 336–--347.
  [Online]. Available: \url{https://ieeexplore.ieee.org/document/9401982/}
\BIBentrySTDinterwordspacing

\bibitem{mcnemar}
Q.~McNemar, ``Note on the sampling error of the difference between correlated
  proportions or percentages,'' \emph{Psychometrika}, vol.~12, no.~2, pp.
  153--157, 1947.

\bibitem{MenziesBCMLSTZ13}
\BIBentryALTinterwordspacing
T.~Menzies, A.~Butcher, D.~R. Cok, A.~Marcus, L.~Layman, F.~Shull, B.~Turhan,
  and T.~Zimmermann, ``Local versus global lessons for defect prediction and
  effort estimation,'' \emph{{IEEE} Trans. Software Eng.}, vol.~39, no.~6, pp.
  822--834, 2013. [Online]. Available:
  \url{https://doi.org/10.1109/TSE.2012.83}
\BIBentrySTDinterwordspacing

\bibitem{morales_repor_2020}
\BIBentryALTinterwordspacing
R.~Morales, F.~Khomh, and G.~Antoniol, ``\BIBforeignlanguage{en}{{RePOR}:
  {Mimicking} humans on refactoring tasks. {Are} we there yet?}''
  \emph{\BIBforeignlanguage{en}{Empirical Software Engineering}}, vol.~25,
  no.~4, pp. 2960--2996, Jul. 2020. [Online]. Available:
  \url{https://doi.org/10.1007/s10664-020-09826-7}
\BIBentrySTDinterwordspacing

\bibitem{DBLP:conf/wcre/OguraMHK18}
\BIBentryALTinterwordspacing
N.~Ogura, S.~Matsumoto, H.~Hata, and S.~Kusumoto, ``Bring your own coding
  style,'' in \emph{25th International Conference on Software Analysis,
  Evolution and Reengineering, {SANER} 2018, Campobasso, Italy, March 20-23,
  2018}, R.~Oliveto, M.~D. Penta, and D.~C. Shepherd, Eds.\hskip 1em plus 0.5em
  minus 0.4em\relax {IEEE} Computer Society, 2018, pp. 527--531. [Online].
  Available: \url{https://doi.org/10.1109/SANER.2018.8330253}
\BIBentrySTDinterwordspacing

\bibitem{codex}
{OpenAI}, ``{OpenAI Codex \url{https://openai.com/blog/openai-codex}}.''

\bibitem{9678712}
M.~Paltenghi and M.~Pradel, ``Thinking like a developer? comparing the
  attention of humans with neural models of code,'' in \emph{2021 36th IEEE/ACM
  International Conference on Automated Software Engineering (ASE)}, 2021, pp.
  867--879.

\bibitem{Pearce_Ahmad_Tan_Dolan-Gavitt_Karri_2021}
\BIBentryALTinterwordspacing
H.~Pearce, B.~Ahmad, B.~Tan, B.~Dolan-Gavitt, and R.~Karri, ``Asleep at the
  keyboard? assessing the security of github copilot’s code contributions,''
  no. arXiv:2108.09293, Dec 2021. [Online]. Available:
  \url{http://arxiv.org/abs/2108.09293}
\BIBentrySTDinterwordspacing

\bibitem{copilotcopyright}
F.~Reda, ``{GitHub Copilot is not infringing your
  copyright~\url{https://felixreda.eu/2021/07/github-copilot-is-not-infringing-your-copyright/}}.''

\bibitem{sobania2023analysis}
D.~Sobania, M.~Briesch, C.~Hanna, and J.~Petke, ``An analysis of the automatic
  bug fixing performance of chatgpt,'' 2023.

\bibitem{copilotLicense}
StephanieGlen, ``{Developers warned: GitHub Copilot code may be
  licensed~\url{https://www.techtarget.com/searchsoftwarequality/news/252526359/Developers-warned-GitHub-Copilot-code-may-be-licensed}}.''

\bibitem{tabnine}
{Tabnine}, ``{AI assistant for software developers
  \url{https://www.tabnine.com/}}.''

\bibitem{wang_bridging_2022}
\BIBentryALTinterwordspacing
D.~Wang, Z.~Jia, S.~Li, Y.~Yu, Y.~Xiong, W.~Dong, and X.~Liao, ``Bridging
  pre-trained models and downstream tasks for source code understanding,'' in
  \emph{Proceedings of the 44th {International} {Conference} on {Software}
  {Engineering}}, ser. {ICSE} '22.\hskip 1em plus 0.5em minus 0.4em\relax New
  York, NY, USA: Association for Computing Machinery, Jul. 2022, pp. 287--298.
  [Online]. Available: \url{https://dl.acm.org/doi/10.1145/3510003.3510062}
\BIBentrySTDinterwordspacing

\bibitem{wang2021codet5}
Y.~Wang, W.~Wang, S.~Joty, and S.~C. Hoi, ``Codet5: Identifier-aware unified
  pre-trained encoder-decoder models for code understanding and generation,''
  \emph{arXiv preprint arXiv:2109.00859}, 2021.

\bibitem{10.1145/3485275}
\BIBentryALTinterwordspacing
C.~Watson, N.~Cooper, D.~N. Palacio, K.~Moran, and D.~Poshyvanyk, ``A
  systematic literature review on the use of deep learning in software
  engineering research,'' \emph{ACM Trans. Softw. Eng. Methodol.}, vol.~31,
  no.~2, mar 2022. [Online]. Available:
  \url{https://doi-org.univaq.idm.oclc.org/10.1145/3485275}
\BIBentrySTDinterwordspacing

\bibitem{ZHANG2022106922}
\BIBentryALTinterwordspacing
F.~Zhang, X.~Yu, J.~Keung, F.~Li, Z.~Xie, Z.~Yang, C.~Ma, and Z.~Zhang,
  ``Improving stack overflow question title generation with copying enhanced
  codebert model and bi-modal information,'' \emph{Information and Software
  Technology}, vol. 148, p. 106922, 2022. [Online]. Available:
  \url{https://www.sciencedirect.com/science/article/pii/S0950584922000763}
\BIBentrySTDinterwordspacing

\end{thebibliography}
